\documentclass[iop,njp,a4paper,preprint,superscriptaddress,showpacs]{revtex4-1}
\usepackage{graphicx,graphics,color,epsfig}
\usepackage{amsmath}
\usepackage{amsfonts}
\usepackage{amssymb}
\usepackage{appendix}

\begin{document}
\title{Two-dimensional group delay in graphene probed by spin precession measurements}

\author{Yu Song}\email{kwungyusung@gmail.com}
\affiliation{Department of Physics and State Key Laboratory of Low-Dimensional
Quantum Physics, Tsinghua University, Beijing 100084, People's Republic of China}

\author{Han-Chun Wu}
\affiliation{CRANN and School of Physics, Trinity College Dublin, Dublin 2, Ireland}

\author{Yong Guo}
\affiliation{Department of Physics and State Key Laboratory of Low-Dimensional
Quantum Physics, Tsinghua University, Beijing 100084, People's Republic of China}

\begin{abstract}
We take graphene as an example to demonstrate that the present widely adopted expression is only the scattering component of a true 2D group delay in the condensed matter context, in which the spatial Goos-H\"{a}nchen (GH) shift along an interface contributes an intrinsic component.
We relate the dwell time to spin precession and derive a relation between the 2D group delay and dwell time, whereby we for the first time reveal that, the group delay for 2D ballistic electronic systems can be directly observed by measuring a conductance difference in a weak-field spin precession experiment. This physical observable not only implies the group delay being a relevant quantity even in the condensed matter context, but also provides an experimental evidence for the intrinsic effect of the GH shift.
Finally, we revisit the 2D Hartman effect, a central issue of the group delay, by analytically solving it via the vested relation and calculating the proposed observable at the Dirac point.
\end{abstract}
\pacs{
03.65.Xp, 	
72.80.Rj, 	
85.75.-d   
}
\date{\today}
\maketitle

\section{Introduction}
The tunneling of a particle through a barrier is one of the most
ubiquitous and fundamental quantum processes.
Over eighty years ago, it was suggested that there is a time duration associated
with such a process \cite{propose}.
Among the various proposed expressions \cite{review,review0},
Wigner-Smith delay or group delay ($\tau_{g}$) \cite{groupdelay,Smith}
and dwell time ($\tau_{d}$) \cite{Smith,dwelltime} are two well-established
times, which describe the reflection or transition of a pulse
peak and the momentary capture and release of a
tunneling particle, respectively.
1D tunneling group delay stems from the phase shift introduced
by scattering at the interfaces \cite{particle}.
It can be given by the eigenvalues of the Wigner-Smith time-delay matrix,
or equivalently expressed by the change rate of the phase shift
with respect to angular frequency \cite{groupdelay,Smith}
\begin{equation}
\tau^{St}_{g}=\sum_\xi|\xi|^2\hbar\frac{d\phi_\xi}{dE},\quad(\xi=r,t)
\end{equation}
where $r=|r|e^{i\phi_r}$ ($t=|t|e^{i\bar{\phi}_t}$) and $\phi_r$
($\phi_t\equiv\bar{\phi}_t+k_xl$, see, Fig. 1b) are the reflection (transmission)
coefficient and corresponding phase shift, respectively.
In the 2D tunneling case,
there generally is a nonzero
lateral Goos-H\"{a}nchen (GH) shift \cite{GH} of the reflected or transmitted beam
along corresponding interface \cite{GHg3,GHsc,GHg1,GHg2,giantGH} (see Fig. 1b).
Obviously, it intrinsically modifies the phase shift at the interface,
thus plays an important role in the 2D group delay.
This fact has been noticed in optics \cite{optics}
but long-term ignored in the condensed matter context.
Instead, Eq. (1) containing only the scattering component has been widely
used in earlier reports
(see, e.g., Refs. \cite{PRL,hartman1,hartman2,precession,ssc,yiyanggong,physB,recent}).
Accordingly, the 2D Hartman effect \cite{hartman}, a central
issue of the group delay,
needs to be revisited.

On the other hand, measurements of the tunneling process are essential
for all attosecond experiments.
Very recently, group delay measurements are eventually achieved in
\emph{atoms and molecules} by approaches with rich skills \cite{measure1,measure2}.
However, these approaches are not feasible in \emph{condensed matters} due to the fact
that the numerous electrons on the Fermi surface are hard to be detected individually.
For a long term, this dilemma has made the group delay being
considered to be of little physical significance in this context \cite{review0}.
So, to propose a feasible approach
in such a context is a fundamental subject and still a key challenge
in the field.
Very recently, the authors in Ref. \cite{precession} made an attempt
in graphene by using a Larmor clock \cite{IBM,van}.
However, the analysis was restricted to the Dirac point
and the improper definition of 2D group delay was adopted.

In this work, we
demonstrate the effect of the GH shift on the 2D group delay
and propose an approach for measuring group delay
in 2D ballistic electronic systems.
We take graphene \cite{graphene1} as an example and show that the GH
shift contributes an intrinsic component which adds an asymmetric feature
to the 2D group delay.
We find that the 2D dwell time can be related to a weak-field spin
precession experiment
and we derive a relation between the group delay and dwell time,
whereby we reveal that, the 2D group delay (and hence the intrinsic effect of
the GH shift on it) can be directly observed through simple conductance measurements
in this experiment.
Such an approach provides a general tool for group delay measurement
in 2D ballistic electronic systems, and is much easier to be realized than the ones used in Refs. \cite{measure1}
and \cite{measure2}.
We at last \emph{analytically} resolve the 2D Hartman effect in graphene
and investigate the proposed observable at the Dirac point.

\section{intrinsic effect of GH shift on 2D group delay}
Let us consider a 2D quantum tunneling through a barrier in graphene \cite{system} (see, Fig. 1).
The electron possesses Fermi energy $E$ and a central incident angle of $\alpha$.
In the stationary state description, the electron can be
represented as a wave packet as a weighted
superposition of plane waves
(each being a solution of Dirac's equation).
The appearing locus and time of the packet peak (equivalently, the
electron) are determined
by the condition that the gradient of the total phase in the wave vector
($\boldsymbol{k}=k_x\hat{x}+k_y\hat{y}$) space must vanish.
This is similar to the optical case \cite{cfli}.
A comparison of the conditions in direction
for the reflected (transmitted) and the incident beam provides
corresponding lateral shift \cite{GHg1,GHg2,giantGH},
$\sigma^\pm_r=-d\phi_r/dk_y\mp1/2k_x$ and $\sigma^\pm_t=-d{\phi}_t/dk_y$ [see Fig. 1(b)].
Here `$\pm$' stands for the $A$ and $B$ component of the graphene spinor, respectively.
A comparison of the conditions in magnitude
gives the average 2D group delay in reflection and
transmission,
$\tau^{\xi}_{g}=\hbar d(\phi_{\xi}+k_y\sigma_{\xi})/dE,$
where average values $\sigma_{\xi}=-d\phi_{\xi}/dk_y$ ($\sigma_r=\sigma_t$
for a symmetric barrier)
have be used since the two components are simply added in the group delay.
For asymmetric barriers, there is a difference between $\tau^r_{g}$ and
$\tau^t_{g}$, and the 2D group delay is defined as
$\tau_{g}=\sum_{\xi}|\xi|^2\tau^{\xi}_{g}$.
In view of Eq. (1), it can be rewritten as
\begin{equation}
\tau_{g}=\tau^{St}_{g}+\tau^{GH}_{g},
\end{equation}
where $\tau^{St}_{g}$ is due to the scattering and
$\tau^{GH}_{g}=\sum_{\xi}|{\xi}|^2\sigma_{\xi}\sin\alpha/v_F$ results from the GH shift.
It is noted that the GH component has been widely ignored in the condensed matter context, see, e.g., Refs. \cite{PRL,hartman1,hartman2,ssc,precession,yiyanggong,physB,recent}.
At time $\hbar d{\phi}_t/dE$ the wave front of the transmitted beam reaches $(l,0)$ and
then propagates freely with a velocity of $v_F$ to the final position
$(l,\sigma_t)$,
which will cost a duration of $\sigma_t\sin\alpha/v_F$ since the wave front is perpendicular to the propagation direction.
This picture also holds for the reflected beam.
A weighted average of the transmission and reflection gives $\tau^{GH}_{g}$.

Fig. 2 shows clearly the contribution of the GH shift to the 2D group
delay.
As is seen, the scattering component is symmetric
about the center of the transmission gap (TG) ($E/V=\cos^{-2}\alpha$)
due to the symmetry of $\phi_{\xi}$ about the center.
Contrarily, the GH component is asymmetric about the TG's
center, which stems from the behavior of the quantum GH shift.
The GH shift has the same trend as the classical
shift ($\sigma_S=l\tan\beta$ with
$\sin\beta=\hbar v_Fk_y/(E-V)$ the refracted angle) predicted by the Snell's law \cite{GHg2},
which is negative (positive) in the low (high) energy region (see Fig. 2).
Accordingly, through the intrinsic term in the phase shift
($\phi_{\xi}+k_y\sigma_{\xi}$), the GH shift not only
quantitatively contributes a part of order of $\sigma_S\sin\alpha/v_F$
in the 2D group delay but also, qualitatively, results in the remarkable
asymmetric feature of the 2D group delay.


\section{measurement of the group delay by spin precession}
We now seek physical observable for the 2D group delay in graphene. 
The Larmor precession of the electron spin in a magnetic field provides
a clock for studying the electron dynamics \cite{Larmor-clock,IBM,precession}.
Here, we calculate the transmission probability ($T=|t|^2$) and $\tau_d$
in a geometry where the magnetic field is applied
in the graphene plane \cite{precession}.
(Note the 2D dwell time has the same definition as the 1D one.)
An explicit equality between the dwell time and an average transmission probability
$<T>\equiv(T_{zy}-T_{\bar{z}y})/
(T_{zy}+T_{\bar{z}y})$ is found in the weak-field limit:
$\omega_B\tau_d=<T>|_{\mathcal{B}\rightarrow0}$ (see, Fig. 3).
Here $\omega_B=g\mu_BB$ is the Larmor frequency, $g$ the gyromagnetic
factor, $\mu_B$ the Bohr magneton, and $B$ the magnetic field.
The weakness of the magnetic field can be described by its reduced
strength, $\mathcal{B}\equiv\hbar\omega_B/2E_0$.
$T_{z(\bar{z})y}$ is the transmission probability for an electron
incident with $y$-directed spin and transmitted with $z(\bar{z})$-directed spin.
This equation holds for \emph{all Fermi energies and incident angles} and can be interpreted physically in following.
Multiplying the right side by $\hbar/2$ gives the expectation
value of $S_z$, which is just determined by the Larmor frequency and the time
the precession persists (obviously, the dwell time rather than
$\tau_g^{St}$ shown in Ref. \cite{precession}) as the left side expressed.
Note $T$ implies an experimental observable, i.e.,
a conductance $G(E)=G_0\int_{-\pi/2}^{\pi/2}T(E,\alpha)d(\sin\alpha)$,
with the conductance quantum of $G_0=2e^2/h$ accounting for a twofold valley degeneracy.
Thus, the above equality can be rewritten as
\begin{equation}
\int_{-\pi/2}^{\pi/2}\tau_d(E,\alpha)T(E,\alpha)d(\sin\alpha)
=\frac{G_{zy}(E)-G_{\bar{z}y}(E)}
{2\omega_BG_0}|_{\mathcal{B}\rightarrow0},
\end{equation}
which clearly indicates that the dwell times can be related
to a spin precession experiment.

We now try to relate the group delay to the conductance by obtaining
the relation between it and the dwell time.
Due to the 2D feature of the tunneling and spinor nature of graphene,
the derivation for the relation is much more skillful and complex than the one for 1D case
of normal Fermions shown in Ref. \cite{particle}.
The detailed derivation can be found in the Appendix, here we
give the concise result
for a rectangular potential barrier 
\begin{equation}\label{graphene}
\tau_{d}=\tau_{g}+\tau_{i},
\end{equation}
where $\tau_{i}=\hbar[\textmd{Re}(r)\cos\alpha+\textmd{Im}(r)\sin\alpha]\sin\alpha/E\cos^2\alpha$
is a self-interference delay stemming from the interference of the incident and
reflection wave functions \cite{review,interference}.
Note this term has so far been widely believed to disappear
\cite{hartman1,hartman2,ssc,yiyanggong}, where the GH component
was ignored and an improper variation method that is simply extended from Ref. \cite{particle} was taken.
Fig. 4 shows the group delay, dwell time, and self-interference
delay in reduced form as a function of the Fermi energy at a
fixed incident angle, where $\epsilon\equiv E/E_0$ is a reduction factor.
The reduced self-interference delay ($-\epsilon \tau_{i}$) achieves the
maximum at the TG's center
and oscillates around zero outside the gap.
The self-interference delay itself is important only in the low energy region 
(diverging as $E^{-1}$ when $E\rightarrow0$), and disappears
at (anti)resonant tunneling since there
is no interference in front of the barrier.
Accordingly, the group delay nearly coincides with the dwell time except within the
low energy ranges or around the TG's center.

The average self-interference delay, $\int_{-\pi/2}^{\pi/2}\tau_iTd(\sin\alpha)$ oscillates
with $E$ and disappears at a relatively high Fermi energy (see the insert in Fig. 4).
Therefore, Eqs. (3) and (4) imply that, for any Fermi energy compared with $V$,
the 2D group delay can now be directly
observed by spin associated conductance difference in a spin precession
experiment
\begin{equation}
\int_{-\pi/2}^{\pi/2}\tau_g(E,\alpha)T(E,\alpha)d(\sin\alpha)
\approx\frac{G_{zy}-G_{\bar{z}y}}
{2\omega_BG_0}|_{\mathcal{B}\rightarrow0}.
\end{equation}
As is seen in Fig. 5, the spin associated conductance difference
which can be obtained by measuring $G_{zy}$ and $G_{\bar{z}y}$ (see insert in Fig. 5)
increases for weaker $\mathcal{B}$.
At $\mathcal{B}=10^{-3}$, it is already a rather good measurement of the
2D group delay for $E>0.2V$.
Thus far, we have provided a feasible and relatively easy approach for
measuring group delay in graphene,
which can be applied to other 2D ballistic electronic systems, such as normal semiconductors,
bilayer and multilayer graphene, surface states of 3D topological insulators,
and their heterojunctions.
(Superconductors will be an exception as superconducting states
may be destroyed by the magnetic fields).
Since the experimental measurement is feasible, the group delay in ballistic electronic systems should
be regarded as a relevant quantity.
Meanwhile, the scattering component is larger (smaller) than the expected value of the group delay
when $E<V$ ($E>V$), a result of the GH shift of different signs.
Then a comparison between the observed value of the conductance difference
(right side of Eq. (5)) and the theoretical prediction of the average group delay
(left side of Eq. (5)) can be utilized to probe the intrinsic effect of the GH shift.

\section{the 2d Hartman effect in graphene: analytical results and physical observable}
We now revisit the 2D Hartman effect in graphene analytically.
We find that the self-interference delay can be
related to the difference in the expectation values of $\hat{p}_x$
at the two outer 
or inner 
boundaries of the barrier,
$\tau_{i}=-\hbar^{2}(\psi^{*}\partial\psi/\partial x)|_{0^\mp}^{l^\pm}/2 j_{in}p_x^{2}$.
This can be rewritten as
$\tau_{i}=q_{x}^{2}\tau_{2(1)}/p_{x}^{2}$ for $q_{x}^{2}>0$ $(q_{x}^{2}<0)$,
where $\tau_{1(2)}=P_{1(2)}/{j_{in}}$, 
$j_{in}=v_F\cos\alpha$ is the flux of incident particles,
and $P_{1(2)}$ is the intensity (interference) component of the
stored probability.
Based on this equation and Eq. (4) (note $\tau_d=\tau_1+\tau_2$)
the group delay can be rewritten as
$\tau_{g}=\tau_{1(2)}+(1-\lambda^2)\tau_{2(1)}/\cos^2\alpha$ for $q_{x}^{2}>0$ $(q_{x}^{2}<0)$,
where $\lambda=(E-V)/E$ is a ratio of the kinetic energies inside
and outside the barrier (see the appendix).
The probability density for the evanescent mode ($\kappa^2\equiv-q_x^2>0$)
simply decays exponentially.
Taking the analytical
expression of $P_{1(2)}$ by the transfer matrix method,
one can find in the limit $l\rightarrow\infty$,
$P_{1}\rightarrow E\cos^{2}\alpha/V\kappa$,
and
$P_{2}\rightarrow0$.
Thus 
the dwell time, the group delay, and the self-interference delay are all saturated in this exponential limit:
$\tau_{d}\rightarrow(\kappa v_{F})^{-1}E\cos\alpha/V$,
$\tau_{g}\rightarrow(\kappa v_{F})^{-1}(1+\lambda)/\cos\alpha$,
and $\tau_{i}\rightarrow(\kappa v_{F})^{-1}[E\cos^2\alpha-(1+\lambda)V]/V\cos\alpha$.
The analytical results obtained here clearly indicate that the 2D Hartman effect
does happen in graphene provided $q_x$ becomes imaginary under the barrier,
which is the same condition for semiconductor-based
structures with parabolic dispersions \cite{particle}.
This result is not only quantitatively different from Ref. \cite{hartman1}
where only the scattering component was considered,
but also essentially opposite to the Dragomans' conclusion that the Hartman effect is absent in graphene
as declaimed in Ref. \cite{hartman2}.

We at last devote ourselves to the barrier-length dependent behavior of
the proposed physical observable.
Since there are no transmission modes, the average group delay is also saturated
with the barrier length at the Dirac point ($E=V$) (see, insert of Fig. 6).
This behavior and the saturation value can be directly detected by
conductance measurements.
Moreover, the group delay is dominated by the scattering component,
which can be well approximated by $E_0\tanh L/E\sin\alpha$
with $L\equiv El\sin\alpha/E_0$.
As can be seen in Fig. 6, although the GH component dramatically increases as $|\alpha|$ and
finally dominates the group delay, it makes negligible contribution
to the average group delay, since the transmission probability
(which can be well approximated by $1/\cosh^2L$)
exponentially decreases with $|\alpha|$.
In other word, the intrinsic effect of the GH shift should be probed
deviating from the Dirac point, which coincides with Fig. 5.
In view of Eq. (5), the saturation value is given by
$(E_0/E)\int _{-\infty }^{\infty }\tanh L/(L\cosh^2L)dL=14\zeta(3)E_0/(\pi^2E).$
For our case, $E/E_0=3\pi$ gives a value of about $0.18$,
which coincides well with the numerical result in Fig. 6.


\section{conclusions and remarks}
In summary, we have demonstrated the intrinsic contribution
of the GH shift to the 2D group delay in graphene.
More importantly, we have proposed a feasible and relatively
easy approach for the observation of group delay (and hence the intrinsic effect of the GH shift)
by conductance measurements in a weak-field spin precession experiment.
We have also analytically revisited the 2D Hartman effect,
and investigated the barrier-length dependent behavior of the proposed observable.
Although quantitative results may vary for other systems which differ in
electronic elementary excitations, the intrinsic
effect of GH shift generally hold for any 2D ballistic electronic
systems, and the proposed conductance difference
provides a universal physical observable for 2D group delay in these systems (except superconductors).
Besides, our results also imply that to construct a generalized Wigner-Smith
time-delay matrix that is valid for 2D case and 3D case is an urgent issue
in the condensed matter physics field.


\section{acknowledgement}
This project was supported by the NSFC (10974109 and 11174168), the SRFDP (20100002110079),
and the 973 Program of China (2011CB606405). 
YS benefited from discussions with C. W. J. Beenakker.
HCW is grateful to the SFI Short Term Travel fellowship support
during his stay at PKU.

\appendix
\section{Derivation for the relation between 2D $\tau_g$ and $\tau_d$ in graphene}

Let us begin with the single-particle Dirac equation that governs the
low-energy excitation in graphene. In the barrier region it reads as
$
[v_{F}\boldsymbol{\sigma}\cdot\mathbf{p}+V(x)]\boldsymbol{\Psi}(x,y)=E\boldsymbol{\Psi}(x,y),
$
where the pseudospin matrix $\boldsymbol{\sigma}$ has components given by
Pauli's matrices and $\mathbf{p}=(p_x,p_y)$ is the momentum operator.
The eigenstates $\boldsymbol{\Psi}(x,y)$ are two-component spinors
with each component being the envelope function
at sublattice site $A/B$ of the graphene sheet.
Due to the translational invariance along the $y$-axis,
the wave function can be separated as $\boldsymbol{\Psi}=[\psi_{A}(x),\,\psi_{B}(x)]^{T}e^{ik_{y}y}$
with $k_{y}=E\sin\alpha/\hbar v_{F}$. The two $x$-part components
are related by a pair of coupling first-order equations
\begin{subequations}\label{coupling}
\begin{equation}
\frac{\partial}{\partial x}\psi_{A}=k_{y}\psi_{A}-\frac{E-V}{i\hbar v_{F}}\psi_{B},
\end{equation}
\begin{equation}
\frac{\partial}{\partial x}\psi_{B}=-k_{y}\psi_{B}-\frac{E-V}{i\hbar v_{F}}\psi_{A},
\end{equation}
\end{subequations}
which implies a decoupled second-order equation for both
the $A$-site and $B$-site components
\begin{equation}\label{decoupling}
\left\{\frac{\partial^{2}}{\partial x^{2}}+(\frac{1}{\hbar v_{F}})^{2}[(E-V){}^{2}-E^{2}\sin^{2}\alpha]\right\}\psi_{i}=0,
\end{equation}
where $\psi_i=\psi_{A,B}$.

We carry out the energy-variational form and conjugate form of Eq.
(A2) and
upon integration over the length of the barrier we find
\begin{equation}
\left.\left(\frac{\partial\psi_{i}}{\partial E}\frac{\partial\psi_{i}^{*}}{\partial x}-\psi_{i}^{*}\frac{\partial^{2}\psi_{i}}{\partial E\partial x}\right)\right|_{x=0}^{x=l}=\int_{0}^{l}\frac{2E\cos^{2}\alpha-2V(x)}{(\hbar v_{F})^{2}}|\psi_{i}|^2dx.
\end{equation}
It is seen that the left (right) part can be related to the group delay
(dwell time),  when is evaluated by the wave function \emph{outside}
(inside) the barrier.
Note Eq. (A3) is only valid \emph{inside} the barrier,
we express the position derivation of the component inside the barrier by
Eq. (A1) and their conjugate form, which then can be
replaced by the corresponding ones outside the barrier.
Then the left part of Eq. (A3) can be evaluated.
For the $A$-component it reads as
$G+F+\cos\alpha (-ir+ir^*)/\hbar v_{F},$
and for the $B$-component, the result becomes
$G+F+\cos\alpha (ie^{-i2\alpha}r-ie^{i2\alpha}r^*)/\hbar v_{F},$
where
$G=\frac{iE}{\hbar v_{F}}\{[B(0)-A(0)]|r|^{2}\phi_r'+[B(l)-A(l)]|t|^{2}\phi_{t}'\},$
$F=\frac{E}{\hbar v_{F}}\{[B(0)-A(0)]|r||r|'+[B(l)-A(l)]|t||t|'\},$
and the relation of lossless barriers $|t|^{2}+|r|^{2}=1$ is used.
The notations are adopted as $O'\equiv \partial O/\partial E$,
$A(x)=\sin\alpha+i\lambda(x)e^{i\alpha}$, $B(x)=\sin\alpha-i\lambda(x)e^{-i\alpha}$,
and $\lambda(x)=\frac{E-V(x)}{E}$, a ratio of the kinetic energy inside and outside the barrier.
Since $\psi^{*}\psi=\psi_{A}^{*}\psi_{A}+\psi_{B}^{*}\psi_{B}$, the
relation for each component should be added to get the relation for the two-component spinor,
which at last reads
\begin{equation}\label{energy}
\begin{split}
&\frac{\int_{0}^{l}[\lambda(x)-\sin^{2}\alpha]|\psi(x)|^{2}dx}{v_{F}\cos\alpha}\\=
&\lambda(0)|r|^2\hbar\phi_r'+\lambda(l)|t|^2\hbar\phi_t'
-i\lambda(0)\hbar|r||r|'-\\
&i\lambda(l)\hbar|t||t|'
+\hbar\frac{[\textmd{Re}(r)\cos\alpha+\textmd{Im}(r)\sin\alpha]\sin\alpha}{k}\frac{\partial k}{\partial E}.
\end{split}
\end{equation}
This is a general result relating the integral weighted probability
density \emph{inside} the
barrier (left part) and the weighted energy-variational behavior \emph{outside}
the barrier (right part). As is seen, the factor $\lambda(x)$ has
a critical role in the relation.

To clearly relate the general result Eq. (\ref{energy}) with both $\tau_d$
and $\tau_g^{St}$, we consider a restricted condition that
$\lambda(x)$ is a constant under the barrier (i.e., a rectangular barrier).
Note this condition is not necessary
for normal semiconductors case \cite{particle}, a reflection of the spinor nature of graphene.
In this case,
the third and fourth terms on the right side of Eq. (\ref{energy})
disappear due to the lossless condition of the barrier
{$|t||t|'+|r||r|'=0$},
and Eq. (\ref{energy}) can be rewritten
in terms of $\tau_d$ and $\tau_g^{St}$, i.e., as a sub-relation reads
\begin{equation}\label{graphene}
\tau_{d}(\lambda-\sin^{2}\alpha)=\tau^{St}_{g}\lambda+\tau_{i}\cos^2\alpha,
\end{equation}
where a self-interference delay is found from the last term of Eq. (A4),
$\tau_{i}=\hbar [\textmd{Re}(r)\cos\alpha+\textmd{Im}(r)\sin\alpha]\sin\alpha/E\cos^2\alpha.$
It is seen that even the GH component is ignored, the ``group delay" in
graphene does not equal the dwell time as indicated in Refs. \cite{hartman1,hartman2,ssc,yiyanggong}.

The angle-variation of Eq. (A2) is straightforward following a
similar way and the variation result gives the sub-relation between
$\tau_d$ and $\tau_g^{GH}$ under the same restricted
condition of constant $\lambda$.
The result reads as
\begin{equation}\label{graphene}
\tau_{d}\sin^{2}\alpha=\tau^{GH}_{g}\lambda+\tau_{i}(\lambda-\cos^2\alpha),
\end{equation}
which, to our best acknowledge, has not been noticed up to now.
Making a simple addition of the sub-relations in Eqs. (A5) and (A6) and
taking into account Eq. (2), we finally get
\begin{equation}\label{graphene}
\tau_{d}=\tau_{g}+\tau_{i}.
\end{equation}
For the normal incident or 1D tunneling case ($\alpha=0$) the
self-interference delay vanishes [see the factor $\sin\alpha$],
since no reflected portion thus no interference happens in front of the
barrier due to the Klein paradox \cite{klein}.

The relation revealed in Eq. (A7) and the expression for
the self-interference delay also hold for the tunneling of
massless Dirac particles in the topological surface states \cite{TI}
where the real electron spin rather than the sublattice structure in
graphene provides the Dirac structure.
It is also an analogue of electromagnetic waves in the photonic
band gap structure, of which the normal incident case was investigated
in Refs. \cite{winful1,winful2}.

\newpage

\begin{figure}
\centering
\includegraphics[width=\linewidth]{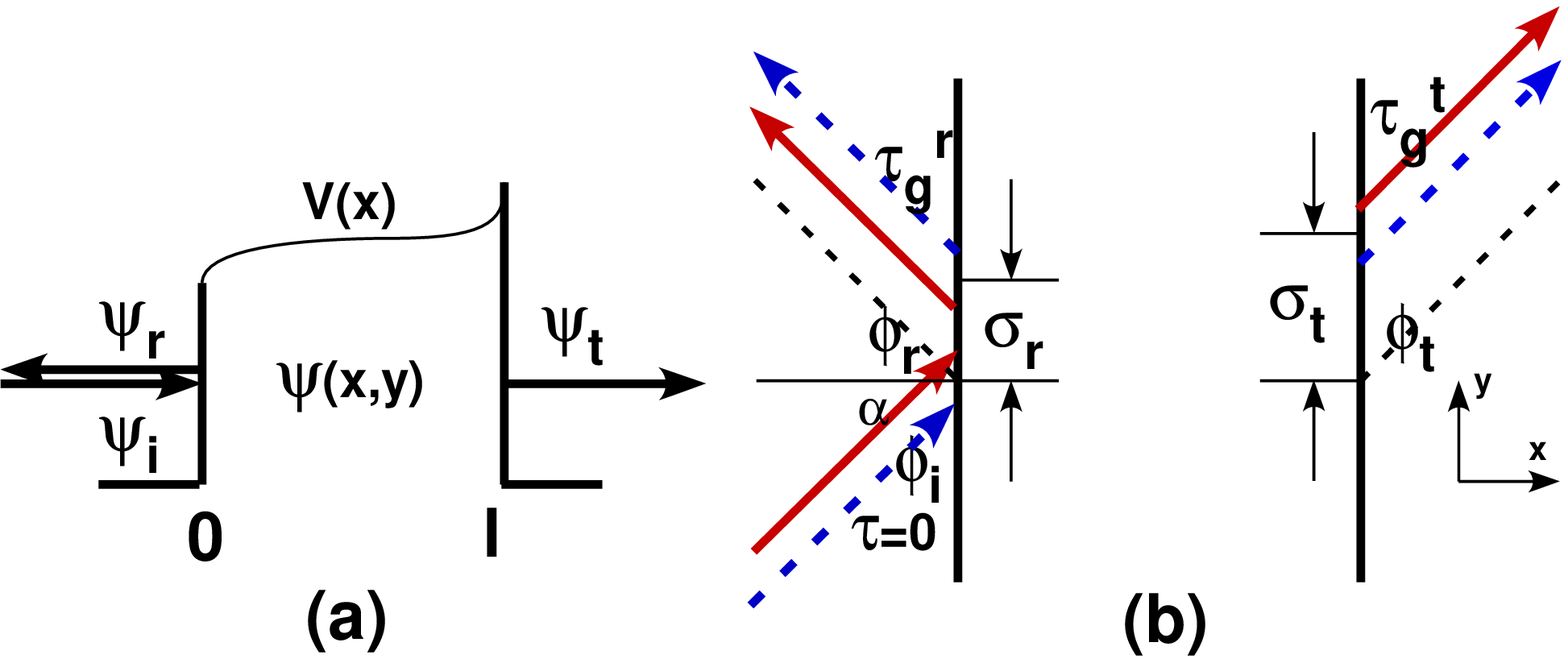}
\caption{
Sectional (a) and top (b) view of schematic diagrams for a particle quantum
tunneling through a potential barrier in graphene.
The red solid and blue dashed lines stand for the trajectories of
the $A$ and $B$ components, respectively.
}
\end{figure}

\begin{figure}
\centering
\includegraphics[width=\linewidth]{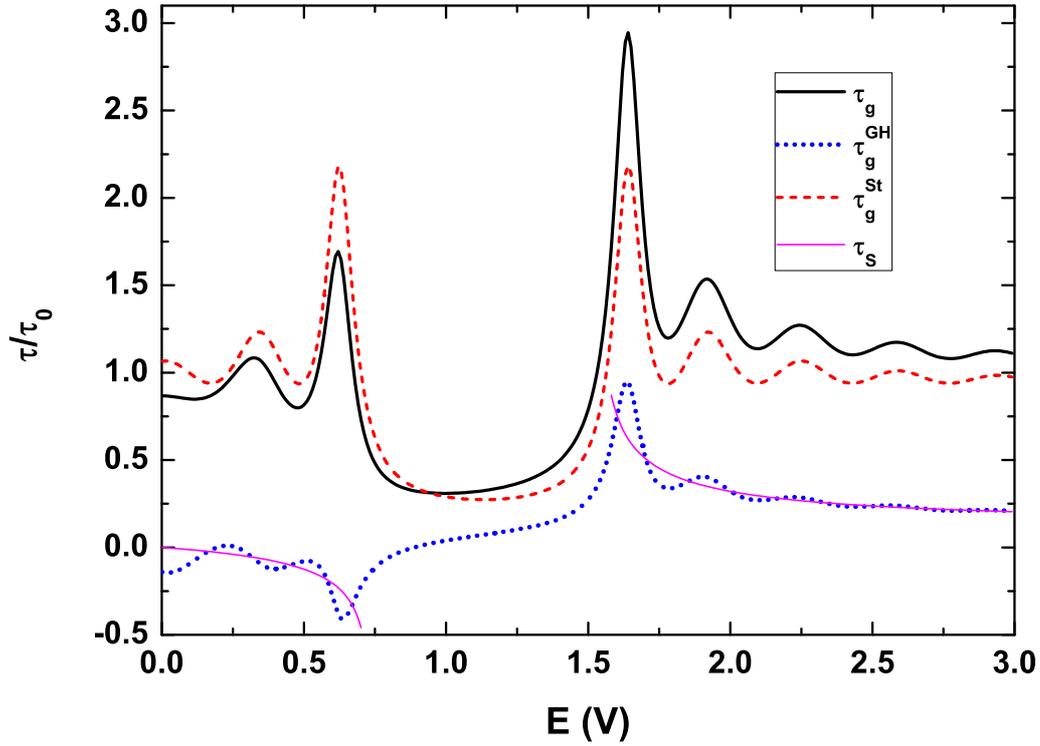}
\caption{
The 2D group delay and its scattering and GH components (in units of the equal
time $\tau_{0}\equiv l/v_{F}$) as a function of the Fermi energy at $\alpha=20^\circ$.
The magenta thin solid curve stands for the time associated with the Snell shift $\tau_S=\sigma_S\sin\alpha/v_F$.
The parameters of the potential barrier (which also hold in Figs. 2-5)
are $l/l_0=1$ and $V/E_0=3\pi$ with $l_0$ and $E_0\equiv\hbar v_F/l_0$ being
a length unit and energy unit, respectively.}
\end{figure}

\begin{figure}
\centering
\includegraphics[width=\linewidth]{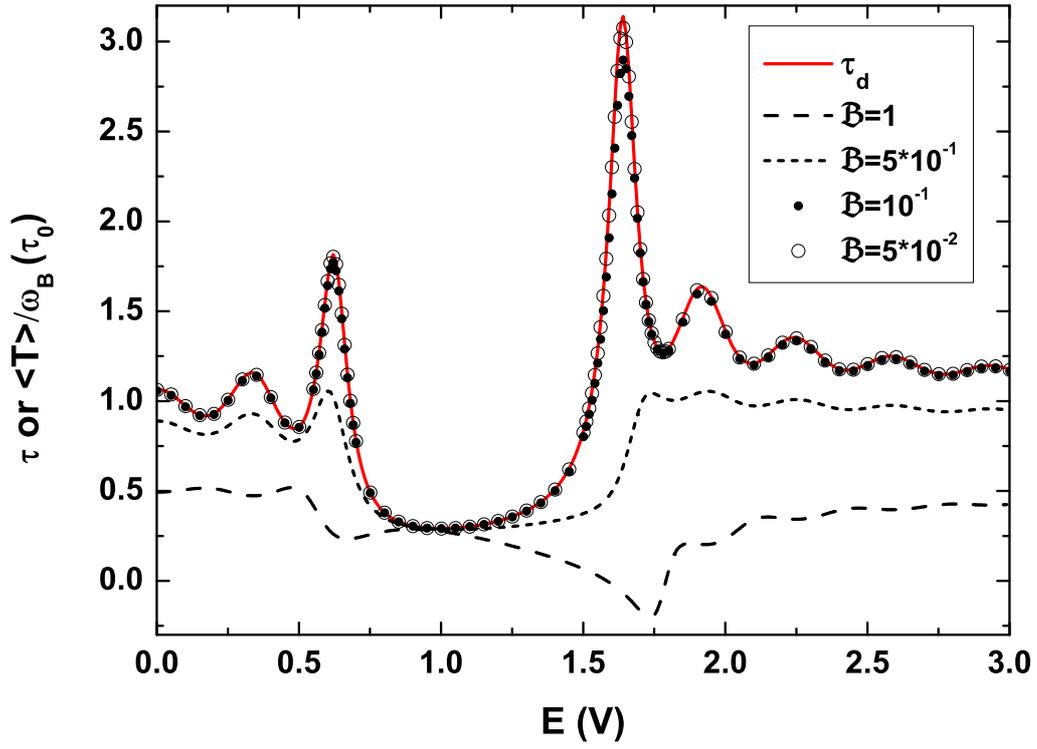}
\caption{
The dwell time and $\mathcal{B}$-dependent $<T>/\omega_B$ as a
function of the Fermi energy at $\alpha=20^\circ$.}
\end{figure}

\begin{figure}
\centering
\includegraphics[width=\linewidth]{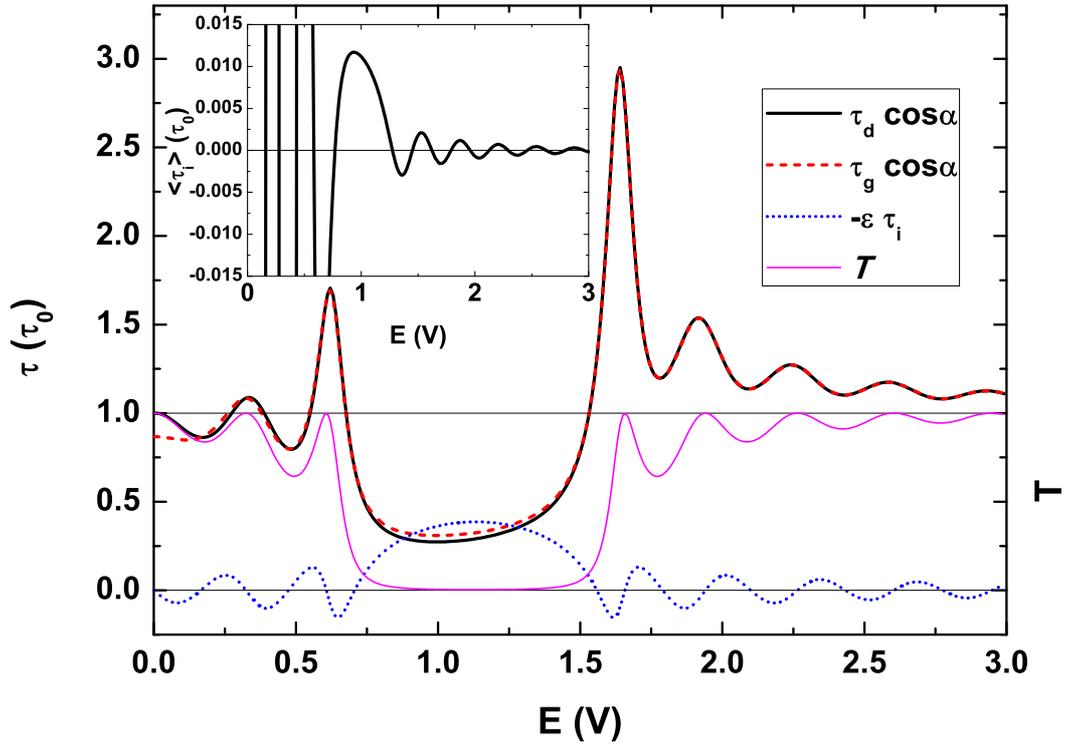}
\caption{
The reduced group delay, dwell time, self-interference
delay, and transmission probability as a function of the Fermi energy at $\alpha=20^\circ$.
Insert: the average self-interference delay versus the Fermi energy.}
\end{figure}

\begin{figure}
\includegraphics[width=\linewidth]{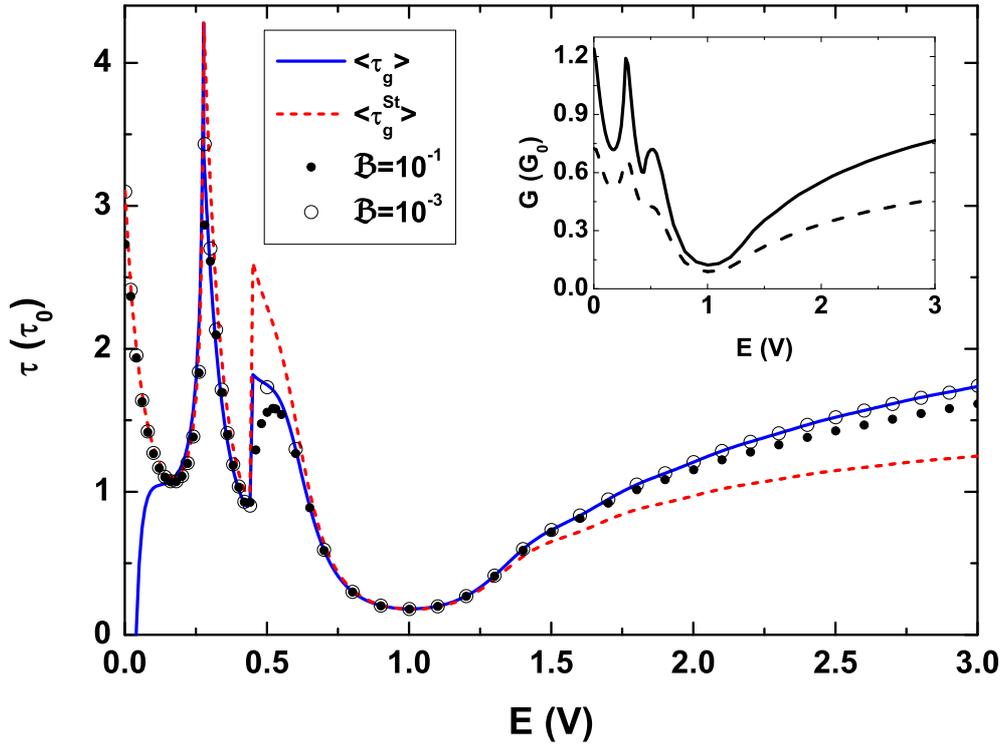}
\caption{
The average group delay, its scattering component, and the $\mathcal{B}$-dependent
conductance difference (i.e., the right hand of Eq. (5))
as a function of the Fermi energy.
Insert: $G_{zy}$ (solid) and $G_{\bar{z}y}$ (dashed) could be directly
measured in the experiment for $\mathcal{B}=10^{-1}$.}
\end{figure}

\begin{figure}
\centering
\includegraphics[width=\linewidth]{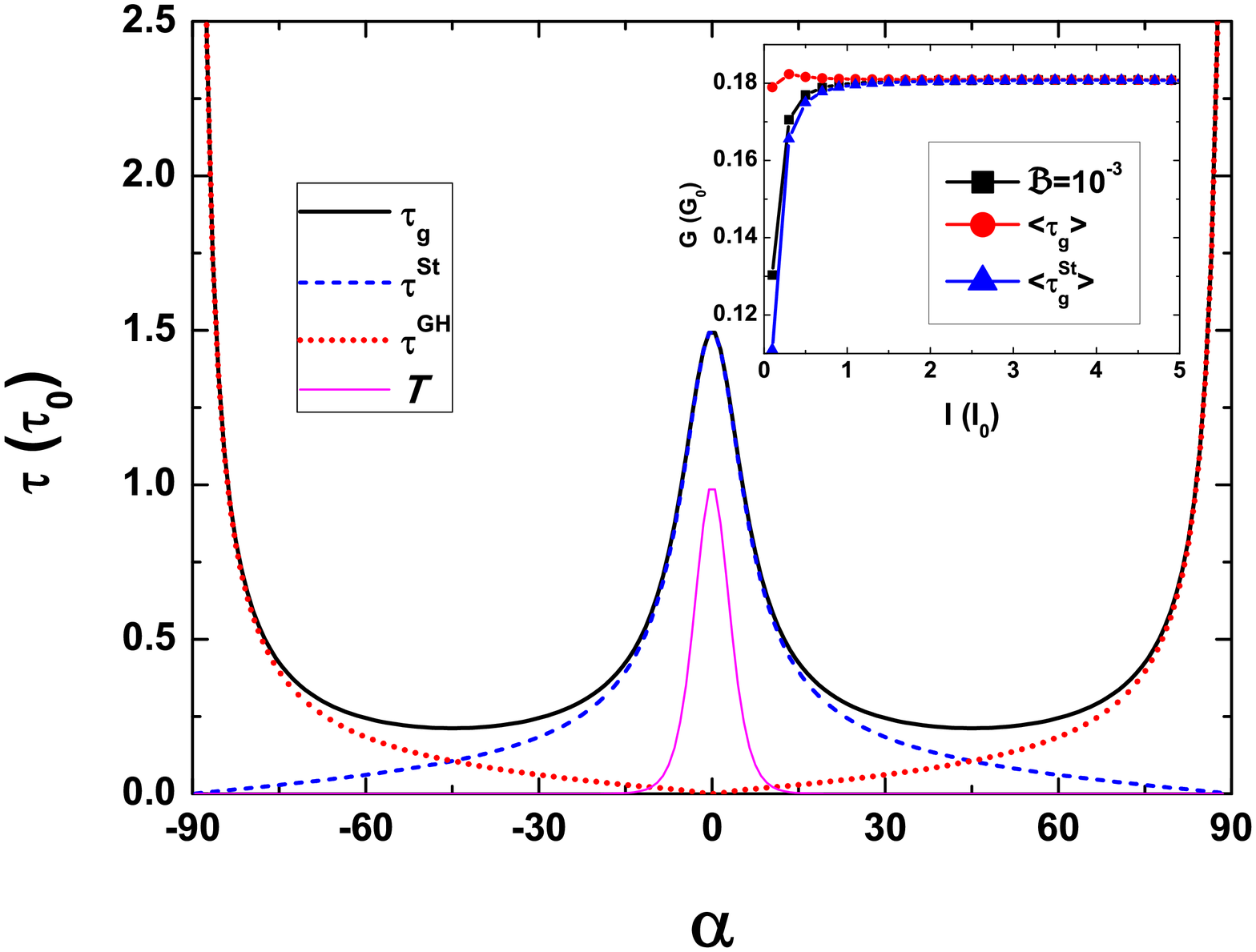}
\caption{
The group delay, its two components, and transmission probability as a function of $\alpha$ at
the Dirac point of a barrier of $V/E_0=3\pi$ and $l/l_0=1.5$.
Insert: the conductance difference, average group delay, and its
scattering component versus the barrier length at the Dirac point for
$V/E_0=3\pi$.}
\end{figure}


\begin{references}
\bibitem{propose}
Condon E U 1931
\emph{Rev. Mod. Phys.} \textbf{3} 43

\bibitem{review}
Hauge E H and St{\o}neng J A 1989
\emph{Rev. Mod. Phys.} \textbf{61} 917

\bibitem{review0}
Landauer R and Martin Th 1994
\emph{Rev. Mod. Phys.} \textbf{66} 217

\bibitem{groupdelay}
Bohm D 1951
\textit{Quantum Theory} (New York: Prentice-Hall);
Wigner E P 1955
\emph{Phys. Rev.} \textbf{98} 145

\bibitem{Smith}
Smith F T 1960
\emph{Phys. Rev.} \textbf{118} 349

\bibitem{dwelltime}
B\"{u}ttiker M 1983
\emph{Phys. Rev.} B \textbf{27} 6178

\bibitem{particle}
Winful H G 2003
\emph{Phys. Rev. Lett.} \textbf{91} 260401


\bibitem{GH}
This shift is named after the physicists who first observed it. See,
Goos F and H\"{a}nchen H 1947
\emph{Ann. Phys.} (Leipzig) \textbf{436} 333

\bibitem{GHsc}
Chen X, Li C F and Ban Y 2006
\emph{Phys. Lett. A} \textbf{354} 161;
Chen X, Lu X J, Wang Y and Li C F 2011
\emph{Phys. Rev.} B \textbf{83} 195409


\bibitem{GHg1}
Beenakker C W J, Sepkhanov R A, Akhmerov A R and Tworzyd\l{}o J 2009
\emph{Phys. Rev. Lett.} \textbf{102} 146804

\bibitem{GHg3}
Wu Z, Zhai F, Peeters F M, Xu H Q and Chang K 2011
\emph{Phys. Rev. Lett.} \textbf{106} 176802

\bibitem{GHg2}
Chen X, Tao J W and Ban Y 2011
\emph{Euro. Phys. J. B} \textbf{79} 203

\bibitem{giantGH}
Song Y, Wu HC and Guo Y 2012
\emph{Appl. Phys. Lett.} \textbf{100} 253116

\bibitem{optics}
Chauvat D, Emile O, Bretenaker F, and Le Floch A 2000
\emph{Phys. Rev. Lett.} \textbf{84} 71


\bibitem{PRL}
P. Pereyra 2000
\emph{Phys. Rev. Lett.} \textbf{84} 1772


\bibitem{precession}
Sepkhanov R A, Medvedyeva M V and Beenakker C W J 2009
\emph{Phys. Rev.} B \textbf{80} 245433

\bibitem{hartman1}
Wu Z, Chang K, Liu J T, Li X J and Chan K S 2009
\emph{J. Appl. Phys.} \textbf{105} 043702

\bibitem{hartman2}
Dragoman D and Dragoman M 2010
\emph{J. Appl. Phys.} \textbf{107} 054306

\bibitem{yiyanggong}
Gong Y and Guo Y 2009
\emph{J. Appl. Phys.} \textbf{106} 064317

\bibitem{ssc}
Esmailpour M, Esmailpour A, Asgari R, Elahi M and Rahimi Tabar M R 2010
\emph{Solid State Commun.} \textbf{150} 655

\bibitem{physB}
Xu X G and Cao J C 2012
\emph{Physica B} \textbf{407} 281

\bibitem{recent}
Sattari F and Faizabadi E 2012
\emph{AIP Adv.} \textbf{2} 012123

\bibitem{hartman}
Hartman T E 1962
\emph{J. Appl. Phys.} \textbf{33} 3427.
For graphene, it has been widely considered in Refs. \cite{hartman1,hartman2,precession}.

\bibitem{measure1}
Eckle P, Pfeiffer A N, Cirelli C, Staudte A, D\"{o}rner R, Muller H G, B\"{u}ttiker M and Keller U 2008
\emph{Science} \textbf{322} 1525

\bibitem{measure2}
Shafir D, Soifer H, Bruner B D, Dagan M, Mairesse Y, Patchkovskii S, Ivanov M Y, Smirnova O and Dudovich N 2012
\emph{Nature} (London) \textbf{485} 343

\bibitem{IBM}
B\"{u}ttiker M 1983
\emph{Phys. Rev.} B \textbf{27} 6178

\bibitem{van}
Tombros N, Jozsa C, Popinciuc M, Jonkman H T and van Wees B J 2007
\emph{Nature} (London) \textbf{448} 571

\bibitem{graphene1}
Novoselov K S, Geim A K, Morozov S V, Jiang D, Zhang Y,
Dubonos S V, Grigorieva I V and Firsov A A 2004
\emph{Science} \textbf{306} 666


\bibitem{system}
The size of the sample is set to be smaller than the electron
mean free path (which is smaller than the coherent length) to ensure
the system stays in the ballistic regime.
The real potential occupies the region of $0<x<l$
and is translational invariant in the $y$-direction.
The sample width in the $y$-direction is rather bigger than $l$
to ensure that the edge detail is not important, see,
Tworzyd J, Trauzettel B, Titov M, Rycerz A and Beenakker C W J,
\emph{Phys. Rev. Lett.} \textbf{96} 246802

\bibitem{cfli}
Steinberg A M and Chiao R Y 1994
\emph{Phys. Rev. A} \textbf{49} 3283


\bibitem{Larmor-clock}
Baz' A I 1967
\emph{J. Sov., Nucl. Phys.} \textbf{4} 182;
\emph{ibid.} \textbf{5} 161;
Rybachenko V F 1967
\emph{J. Sov., Nucl. Phys.} \textbf{5} 635



\bibitem{interference}
Hauge E H, Falck J P and Fjeldly T A 1987
\emph{Phys. Rev.} B \textbf{36} 4203

\bibitem{klein}
Katsnelson M I, Novoselov K S and Geim A K 2006
\emph{Nature Physics} \textbf{2} 620

\bibitem{TI}
Hasan M Z and Kane C L K 2010
\emph{Rev. Mod. Phys.} \textbf{82} 3045;
Qi X L and Zhang S C 2011
\emph{Rev. Mod. Phys.} \textbf{83} 1057

\bibitem{winful1}
Winful H G 2002
\emph{Opt. Express} \textbf{10} 1491

\bibitem{winful2}
Winful H G 2003
\emph{Phys. Rev. Lett.} \textbf{90} 023901


\end{references}
\end{document}